# What Values Do ImageNet-trained Classifiers Enact?

Will Penman[†]
Princeton University
Princeton, NJ, USA
wpenman@princeton.edu

Joshua Babu
Princeton University
Princeton, NJ, USA
jbabu@princeton.edu

Abhinaya Raghunathan
Princeton University
Princeton, NJ, USA
abhinaya@princeton.edu

## ABSTRACT

We identify "values" as actions that classifiers take that speak to open questions of significant social concern. Investigating a classifier's values builds on studies of social bias that uncover how classifiers participate in social processes beyond their creators' forethought. In our case, this participation involves what counts as nutritious, what it means to be modest, and more. Unlike AI social bias, however, a classifier's values are not necessarily morally loathsome. Attending to image classifiers' values can facilitate public debate and introspection about the future of society. To substantiate these claims, we report on an extensive examination of both ImageNet training/validation data and ImageNet-trained classifiers with custom testing data. We identify perceptual decision boundaries in 118 categories that address open questions in society, and through quantitative testing of rival datasets we find that ImageNet-trained classifiers enact at least 7 values through their perceptual decisions. To contextualize these results, we develop a conceptual framework that integrates values, social bias, and accuracy, and we describe a rhetorical method for identifying how context affects the values that a classifier enacts. We also discover that classifier performance does not straightforwardly reflect the proportions of subgroups in a training set. Our findings bring a rich sense of the social world to ML researchers that can be applied to other domains beyond computer vision.

## 1 Introduction

Image classification is a flashpoint for the possibilities and perils of AI broadly. Image classification kicked off convolutional neural networks historically [60][61] and in homage, image classification is used today in introductory exercises for aspiring ML practitioners. Image classification was again a showcase for deep learning [34] with the advent of ImageNet [21]. (By "ImageNet" throughout we refer to the subset of ImageNet stabilized in 2012 in the ImageNet Large Scale Visual Recognition [84], which consists of ~1.2M training images and 50,000 public validation images across 1,000 object categories.) Each year of the ImageNet competition brought new advanced techniques [28] that are now available for a range of AI applications [54]. Today, classifiers boast "better than human" classification accuracy on ImageNet [41]. Image classification itself is considered solved, and the ImageNet contest has now been retired to a Kaggle competition [28]. Researchers now pursue more complex computer vision tasks like object segmentation, scene understanding, question answering, and video-based object detection [28][62]. Thus, image recognition has functioned as a representative of AI perception more broadly, highlighting a steady march toward AI visual understanding of the world.

Precisely because of its ubiquity and power, image classification has recently come under scrutiny, particularly with regard to facial recognition [10]. The discovery that several commercial classifiers are socially biased against dark-skinned women has led to public chastening of AI developers generally [22], healthy calls for regulation [94], and extensions to other object categories [19]. These studies of social bias in image classification have shown that AI is not necessarily a neutral, pure agent that simply goes about the world objectively. AI can be a locus in which social behaviors (including patterns of discrimination) are replicated. Classifiers have the potential to exert unjust social force in high-stakes areas like surveillance [10] and medicine [29] as well as mundane areas like search [70].

We contribute to these developments by showing how image classifiers are always enmeshed in social behaviors. They may not always be socially biased, but they are always involved in making social meaning. We use the language of "values" to name an in-between area that is not about accuracy and not about social bias, but is about the social choices in everyday life. Understanding a classifier's values encourages public debate and interdisciplinary discussion about the future. For instance, values are enacted when YouTube's recommendation system pushes radical content [81], when content moderation tools classify text as "bullying" [15], and when speech systems take on a male or female voice [14]. All of these decisions have significant consequences, but it's an open question socially whether these are the AI decisions we want to make. For image classifiers specifically, a prototypical situation for values to emerge would be in detecting fetuses. Are sonograms and pregnant women's stomachs recognized as images of a "human"? In the US, this is a controversial question that a "human-detector" could not avoid answering. Understanding values also cautions us not to use AI as though it were neutral – a human-detector may be no more "right" in its answer on fetuses than the rest of us are. Overall, because AI choices always involve exclusion of some kind, we need to be mindful of how AI includes and excludes. Understanding values means asking: What ways of living does this AI adopt? Which does it reject?

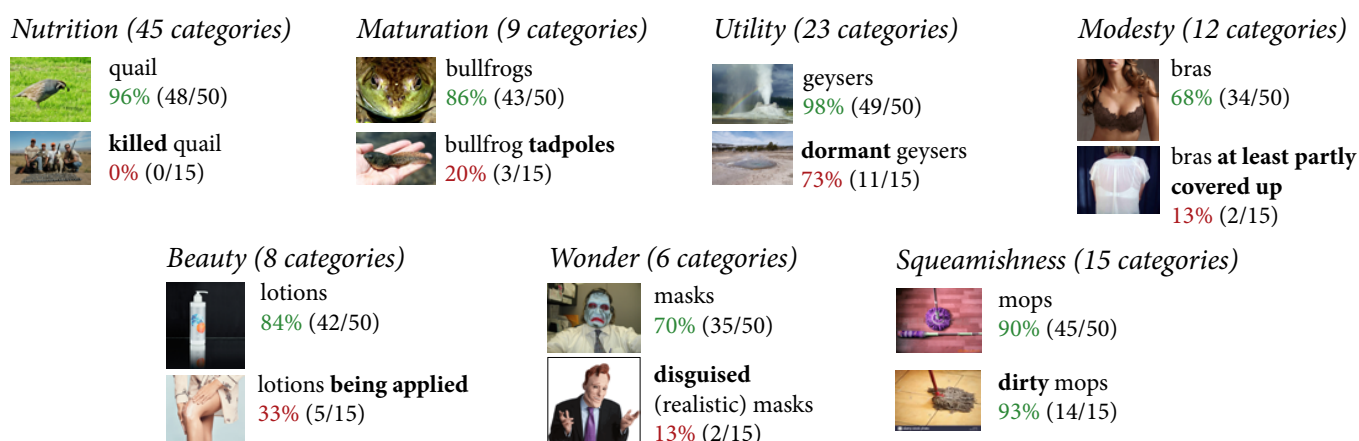

Figure 1. Seven areas in which decision boundaries in ImageNet classifications enact values. Each area represents an open social question without a clear right or wrong answer: what is good to eat? (nutrition); when is something mature enough to be recognized as itself? (maturation); do dormant states count as that thing? (utility); should undergarments be perceived when they're covered up? (modest viewing); do beauty products become part of the person when applied? (beauty); should we notice well-disguised objects? (wonder); and do gross states make something unrecognizable? (squeamishness). Across 118 ImageNet categories, we find that ImageNet-trained classifiers are pescatarians, adult-oriented for living categories, somewhat utility-oreinted, confusingly modest, affirming of beauty efforts, mixed in wonder/mechanism, and not squeamish. We could easily imagine different versions of ImageNet that would train classifiers to value different things and would be equally accurate, but no version would be value-free. Classification results of sample categories are reported from VGG-16 top-5; all differences are significant at p<0.01 or below. Categories are assessed generally on the category's provided 50-image validation set; the value-revealing subgroup is assembled through web searches and assessed as a rival dataset of 15 images (see 3.3).

Our goal in this paper is to identify the values that ImageNet-trained classifiers enact (Fig. 1). The categories and value areas that we examine speak to controversies that are especially relevant for commercial and educational uses of AI.

## 2. Related research

We use Buolamwini & Gebru's [10] foundational research as our point of departure theoretically and methodologically. They found that commercial face classifiers exhibit significant performance disparities across skin-color and gender, especially when analyzed intersectionally (e.g. one classifier had 34.7% error on dark-skinned females despite 12.1% overall error).

### 2.1 Social bias

At a theoretical level, much of the power of Buolamwini & Gebru's [10] analysis comes from implicitly creating a link between how face classifiers are being used (namely, for surveillance) and the social impact that a skin-color/gender performance disparity would create in aggregate given the use (namely, perpetuating racist forms of policing). In this case, because racism is particularly embedded in US society [39][53] and in the West [65], face classifiers that have racial/gender performance disparities can themselves be seen as agents of social impact. This can be called "social bias" and differs from scene bias (including selection bias, capture bias, category/label bias, and negative set bias [98]) in that social bias hurts people and is morally opprobrious. Broadly, then, Buolamwini & Gebru [10] show that when conducting an abstract analysis of a classifier's performance, researchers can make claims about significance by pointing to the classifier's usage. In fact, a proportional analysis is often crucial for observing social bias because AI inequality operates at the level of population and can thereby be difficult to show from individual classification decisions. Moreover, training/testing data is shown to be an effective point of intervention for the systemic causes of AI social bias.

Beyond faces and skin color, others have observed social bias in the form of cultural and income bias in ImageNet categories, in which categories privilege wealthy and Western forms of those categories [19][24].

### 2.2 Perception as a locus for enacting values

We seek to contextualize the social bias line of research without diminishing the importance of identifying, anticipating, and remedying social bias in AI. Given its intellectual origins in remediating representational injustice, research into AI's social biases has treated data proportionality only as problems to be overcome. The social bias research agenda seeks "balanced" [10] datasets that lead to "equal" performance across groups. In this way social bias might be bridled, transcended, subdued [55]. Even the metaphor of a data "audit" [79] implies correctness.

However, research in the humanities and social sciences suggests that human activity goes beyond bias to include values – areas of cultural ambiguity. For scholars of rhetoric, values (and closely related concepts like norms, assumptions, rhetorics, etc.) indicate the need for public debate [4]. In line with philosopher John Dewey [30] values emphasize that the world is a place for discovery and meaning-making. Moving through the world is thereby seen as a value-laden process of taking provisional and contextual answers to questions about the structure and goals of our world, rather than final and objective answers. These answers are not just symbolic, but often material and enacted.

In particular, rhetoric scholars are concerned with how questions go from open to closed and vice versa. How are alt-right ideas making a resurgence in the US? How has the marriage equality movement found popular and legal recognition? Rather than search for timeless, universal claims, these studies are especially tuned to how context, audience, identity, and situation play a role, even in science [59][100]. These highlight disagreement and show how important it is to engage reflectively [17] and collectively [40][85], with special regard for those who have been marginalized [101]. From this perspective, users of AI are not just consumers, but are also citizens and even possible collaborators.

At the level of perception, research in psychology since the 1940s has increasingly shown how object classification is value-laden. In an early experiment, US children overestimated the size of coins more when the coins were worth more (and didn't for comparable cardboard disks), showing that size perception depends on something's *social value* [9]. Then researchers found that across 15 societies, Westerners were very susceptible to certain optical illusions involving perspective drawings whereas some foraging African groups were not fooled at all, showing that we perceive lengths differently, likely based on whether we grow up in a *carpentered physical environment* or not [44][87].

People's attention to objects in a visual field differs by culture; among other findings, Westerners focus on focal objects when describing them, while Japanese people describe scenes more holistically [68] This can be traced to different *metaphysical, epistemological, and social organization/practices assumptions* [68]. Similarly, Americans favor rule-based categorization, in which every item of a group shares at least one feature, but people from Asia categorize images based on family resemblance [69]. Americans easily discard contextual information when reproducing part of an image, and Japanese people easily include it [56].

Beyond culture at large, people's *ability to take action on something* impacts what they perceive, including tennis players who perceive tennis nets as lower when they are able to return more balls and people with a conductor's baton who perceive distances that would've been just out of reach as shorter than people who don't have a tool would guess [104]. People's *languages* also influence how quickly people categorize colors, such as greens and blues [7][80]. Finally, *group identity* impacts what people perceive, including the race of faces and even the physical distance to out-group members [106].

Importantly, these variations are not errors per se [106]; they are evolutionarily and socially beneficial ways of perceiving given a certain context [35].

This project brings rhetoric, psychology, and philosophy to bear on AI perception. Because these cultural ways of seeing are learned largely by age 11 and are solidified by age 20 [43], training/testing datasets for object classifiers function as a formative period, or "ontogeny," for AI, with different algorithms functioning as different kinds of perceptual "formation." Training/testing datasets then also become stabilized windows into the ways of life that the classifier can be expected to join. Depending on the dataset and algorithm, then, classifiers may be adopting perceptual values associated with western, educated, industrialized, rich, and democratic cultures, which are frequently outliers [44].

## 2.3 Integrating accuracy, social bias, and values

We propose a framework for integrating values, social bias, and accuracy. Consider the space $A$ of an agent's possible actions at time $t$. For an ImageNet-trained classifier, $A$ includes all possible top-5 outputs with all possible confidence scores, as well as nothing (if it was processing, or if no image had been presented to it). Now, consider the agent's "world" as the space $W$, which maps possible actions in $A$ to social questions that those actions would speak to. Points in $W$, then, are answers to social questions that a certain set of actions in $A$ would speak to. Further, consider each social question in $W$ as more or less "open" or "closed." Then an agent's "value system" is its locations in $W$ for some range before, during, and after $t$; its answers to relatively open questions are "values"; its answers to relatively closed questions are "correct" or "incorrect"; and its answers to relatively closed questions that deal with people are additionally "social biases."

Consider further that $W$ varies with respect to spaces $C$ (culture) and $R$ (relationality). That is, an agent's world and value system are fuzzy and contested multiple times over: at an individual level within a cultural group, different people think a given question is open, somewhat closed, or closed; at a group level, similarly; and at both levels there are often different mappings from action to social meaning depending on the relationship with an agent (relationality). We call $W'$ (self) an agent's own model of $W$, and note that no AI network we are aware of learns $W'$. Indeed, by some accounts this would involve some level of consciousness [42], and could be viewed as a path for pursuing AGI. With or without agents that learn $W'$, this framework suggests that AI developers need to model $W$ themselves as part of enacting values reflectively.

This model can accommodate cases of correctness and social bias, as well as values. For instance, when classifying an image of a dog, all of $A$ might be mapped to $W$ as correct/incorrect (and maybe partially correct, for other dog breeds) answers to the closed question of "Is this what a giant schnauzer looks like?" We the authors don't know of any giant schnauzer classification controversies that would make this mapping vary much across $C$.

However, *W* would be quite different for parts of *R* that describe anything other than a distant relationship – then, labels like "the dog in the park," "Susan's dog," or "Pepper" would be appropriate answers to the different (but still closed) question of "Which dog is that?" and ImageNet's output would be insufficient. To use a different example, classifying an image of a black person as "gorilla" [88] would be a flagrant case of social bias, an inappropriate answer in *W* to closed social questions about racial inequality/exclusion. In contrast, what we call values deal with locations in *W* that contextually speak to more open social questions about what is good to eat, beautiful, etc. Scholars have found that culture is enacted, contested, and resisted at these edge cases, which have been theorized as the "liminal" or "boundary" [25], the "border" [20], the "threshold" [64], and the "precarious" [45]. These value decisions are happening even if we resolve representational social biases.

Given this framework, the primary task for the rest of this paper is to identify the open questions that ImageNet-trained classifiers' perceptual actions speak to. In line with the assumptions of our approach, we make no claim that the open questions and corresponding values that we identify are objective and timeless [73]; rather, the values that we identify in ImageNet-trained classifiers are argued to hold for a restricted but widespread set of circumstances. We have in mind a broadly American/European cultural milieu (*C*), an impersonal agent-object relationality assumed in most ML work (*R*), and a *t* in the present day. Interpreting decision boundaries as value-laden is a contestable process; that is, these are not themselves objective mappings or even universally held. As an example, elephants and quail can both be hunted, but dead elephants aren't eaten (in today's American/European context), so although we interpret killed quail to visually speak to what is nutritious, if ImageNet recognized killed elephants it would speak to something else.

## 2.4 Data Proportionality Hypothesis

Buolamwini & Gebru [10] also provide methodologically related guidance by taking proportionality to be a rough indicator of prototypicality. They motivate their study by observing that two existing face benchmarks are quite skewed toward lighter skinned people. This enables them to posit that a benchmark "balanced" by gender and skin type would reveal commercial face classifiers' performance disparities across gender and skin type subgroups. In other words, if classifiers are being trained and tested on disproportionate datasets, then we might develop a confident but misleading sense of their abilities. This link from training/testing sets to performance has a commonsense appeal that we might formalize as the Data Proportionality Hypothesis:

**Data Proportionality Hypothesis (DPH)**: Subgroups' proportional representation in training/testing sets predicts classifiers' relative performance on those subgroups.

**Corollary 1**: Increasing training dataset size doesn't necessarily change classifiers' differential performance for subgroups, because it doesn't necessarily change the subgroups' proportional representation.

**Corollary 2**: Overall accuracy rates can hide significant performance disparities among subgroups, especially intersectionally.

The DPH is important for Buolamwini & Gebru [10] because it functions as a heuristic for ML developers to anticipate and intervene in potentially biased classifiers. Moreover, the two corollaries helpfully debunk scale and overall accuracy (respectively) as unfailing remedies for machine learning social bias challenges. However, the DPH is quite loose, with many unaddressed questions. How strongly does the DPH hold in categories beyond faces? [57] How strongly does the DPH hold with affirmative evidence from a training set, rather than just negative evidence from the testing set? And how strongly does the DPH hold when a category subgroup doesn't have an intuitive partition (e.g. male/female)? By answering these questions across more than 100 categories, we give insight into deliberately creating classifiers that enact certain values.

## 3. Method: Rhetorical Values Analysis

Our method for identifying a classifier's values consists of three steps (Fig. 2). Two qualitative analytic steps estimate *W* by generating a mapping from the decision boundary of a category to value. These steps are most useful in exploratory stages, to draw out insights from humanities research and investigate the DPH theoretically. Then a quantitative step assesses where in *W* classifiers act; these are statistically significant assessments of classifiers' enacted values.

**1. Variational overview of category's validation set.** Categories in ImageNet are designed to have variable appearances, positions, view points, poses, background clutter, and occlusions [21]. During training, images are often augmented to increase variability by using random extracted patches and reflecting images horizontally [58], as well as more elaborate data augmentation schemes [16]. Objects themselves are also easier and harder to detect: in competition ImageNet, objects are easier to detect when they occupy most of the image, have a large real-world size, are natural rather than human-made, and have at least some texture [84].

We build on these efforts to understand ImageNet variety. To characterize a category's validation set, we use a heuristic adapted from the field of visual rhetoric [82]. This took the form of writing a short prose paragraph (see parachute example in Fig. 2). This heuristic emphasizes social context, making it easy to generate hypotheses about socially significant decision boundaries. Incidentally, it also help generate hypotheses about social bias, which given the framework presented in 2.3 are understood as confounding factors for understanding a dataset's values.

What Worldview Do ImageNet-trained Classifiers Have?

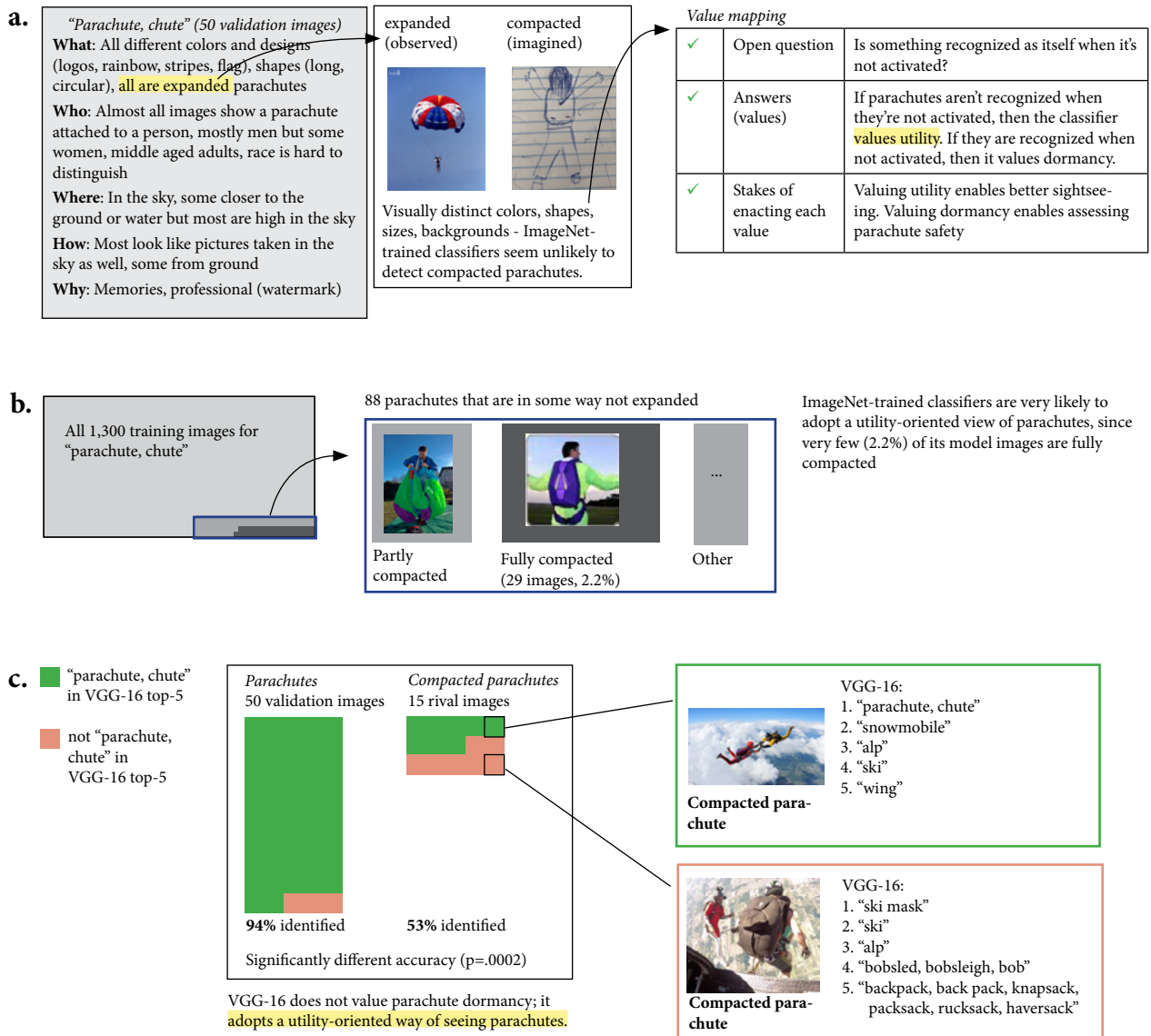

Figure 2. Our method of rhetorical values analysis, shown with the category "parachute, chute." a. A variational overview involves characterizing the category's validation set with a heuristic, inferring a visually distinct presence/absence, and creating a mapping from a potential decision boundary to some value. b. Identifying a category's exceptions from the training set involves obtaining count (as % of training set) through manual inspection and narrowing when desired to sharpen value contrast. Strong hypotheses about classifier performance can now be made. c. Category hypothesis testing involves creating a rival set of value-exposing images, testing comparative accuracy, and interpreting in light of value mapping.

The most delicate and crucial part of a variational overview is developing a mapping from potential decision boundaries to a social value. Translating from decision boundary to values requires identifying an open question of social significance that plays out at the level of object recognition. In the case of parachutes, we suggest that a relevant open question is whether something should be recognized as itself when it's not activated. Open parachutes are very active, while compacted parachutes are very inactive/dormant. We consider this an open question because we can imagine arguments for both perception styles. On one hand, a "parachute-detector" could be desirable for tourists for sightseeing, in which case it would be useful to value utility by not being on the lookout for compacted parachutes. On the other hand, a parachute-detector could be desirable for safety inspectors involved in pre-flight checks, which would require valuing parachute dormancy. Because we elicit contrasting appeals related to significant public concerns, we call it a "rhetorical" values analysis.

We emphasize that, inherent to our definition of value, it is neither inaccurate nor socially biased for a classifier to adopt one value or another. We are not replicating the move to call for a balanced dataset. Thus, we could create a parachute dataset that was "balanced" for active/dormant states, but it would *not be less value-laden*. Assuming classifiers would learn dormant parachutes, that set would simply value object dormancy.

**2. Identification of category's training set exceptions.** We applied the insights from variational analyses by then identifying exceptions from that category's training set. For most categories in ImageNet, that involved manually examining 1,300 images. At this scale, 13 images form only 1% of a dataset's conception of that object, and a single image forms .08% of the whole category.

Examining the training set for a category generates strong evidence for classifiers' performance, because a classifier's performance on a chosen subgroup can be directly traced to its exposure to those during training. Moreover, in the case of training sets like ImageNet that are used as a benchmark for neural architectures, examining the training set also allows us to compare how different classifiers take up values from the same training set (see Section 5). Finally, quantifying exceptions is itself a fuzzy process, and gathering them shows a wide range of variations (marked "other" in Fig. 2c). This reveals shades of difference that the binary statistics in step 3 don't capture well. Identifying training exceptions keeps the doors open to more nuanced testing; our comparisons are meant to elicit only the barest and simplest values.

**3. Category hypothesis testing.** Finally, we assessed ImageNet-trained classifiers' enacted values. To assess classifiers' enacted values, we assembled a set of usually 15 images that functioned as a "rival" [32], or provocative alternative, to what the classifier might be used to seeing (Appendix 3). We searched online for rival images, intentionally mimicking ImageNet's creation process [21], seeking for images to be as varied as the validation set except for the value-revealing point of focus. For collapsed parachutes, that meant having many images taken from the sky of someone in the sky, with a mix of non-professional and professional images. By comparing classifiers' accuracy on the rival validation set to their accuracy on the actual validation set, we can determine with statistical precision whether the rival subgroup is "part of" the object's detectable range of variation. If the rival set scored significantly lower ($p<.01$, often much lower) than the validation set, then we say the classifier "doesn't recognize" images of the rival type. If the rival set accuracy showed some similarity ($p>.1$) with the validation set accuracy, then we say the classifier "recognizes" images of the rival type. To compare accuracy rates, we used Fisher's exact test for significance, which is appropriate given our 2x2 design. At the level of values, the rival set speaks to a potentially different but equally "accurate" sense of that category.

## 4. VGG-16's value system

We conducted variational overviews of 128 ImageNet categories, which involved characterizing in prose each category's 50-image validation set. Consultation with philosophy research suggested that it would be productive to identify open questions in ImageNet categories that are "shape shifters," i.e. have notably indeterminate states. Ships [86] and flowing rivers [38] were examined as shape-shifters by Ancient Greek philosophers. In the mid-20$^{th}$ century, scholars argued that tools sometimes shape-shift, momentarily becoming invisible to us [42], and that in some cases people perceive tools as part of their body [13]. Beyond shape-shifting categories, we estimate that another 65 ImageNet categories (as well as the distribution of categories itself) could be fruitfully analyzed for other values. We leave these for future research.

We then identified training set exceptions for each of the 128 categories, which involved manual inspection of 165,059 images. With this work grounding our hypotheses, we finally gathered over 2,100 rival images across 118 categories and compared classification results. We report on seven value areas adopted by VGG-16 (Appendix 1), but since other classifiers performed similarly (Section 5), we refer to "ImageNet-trained classifiers." Each of these value areas play out in personal, commercial and/or educational classifiers.

Overall, we found moderate coherence across categories. Since humans are in the loop for all of the sources of ImageNet-trained classifiers' value system [33], we expected multiple categories in an area to enact the same value, while also expecting that their value systems would be somewhat strange, having been trained solely from images that were online in 2011 [98].

### 4.1 Nutrition

Nutrition is an open societal question; people's dietary choices can be personal, cultural, geographic, technological, and religious [37][52] and these value differences are reflected in people's identity [49] and even at the level of food terms.[1] Instantiating nutrition is big business; at $1.053 trillion, agriculture, food, and related industries contribute 5.4% of the US GDP [2] The weight loss industry alone is worth $72 billion [99]. Food products are also interesting because they are extreme shape-shifters, going from being alive to being part of another living being.

---

[1] For instance, while useful in day-to-day contexts, the popular idea of a "vegetable" is scientifically incoherent [78]. Specifically, cabbage, broccoli, and cauliflower (along with non-ImageNet categories brussel sprouts, kohlrabi, and kale) are all the same species, but are simply cultivated for different parts [8]; cardoon and artichoke are both products of the same wild progenitor [92]; and although "mushroom" refers to just the fruiting body of a fungus, the other mushroom types in ImageNet refer to the whole fungus. Likewise, using "pork" and "beef" as state-dependent words for food-ready pig and cow has roots in class-based instantiations of nutrition: in the 1000s, the conquering French had the means to eat meat, so the French word for those (*boeuf*, *porc*) came into English [66].

What Worldview Do ImageNet-trained Classifiers Have?

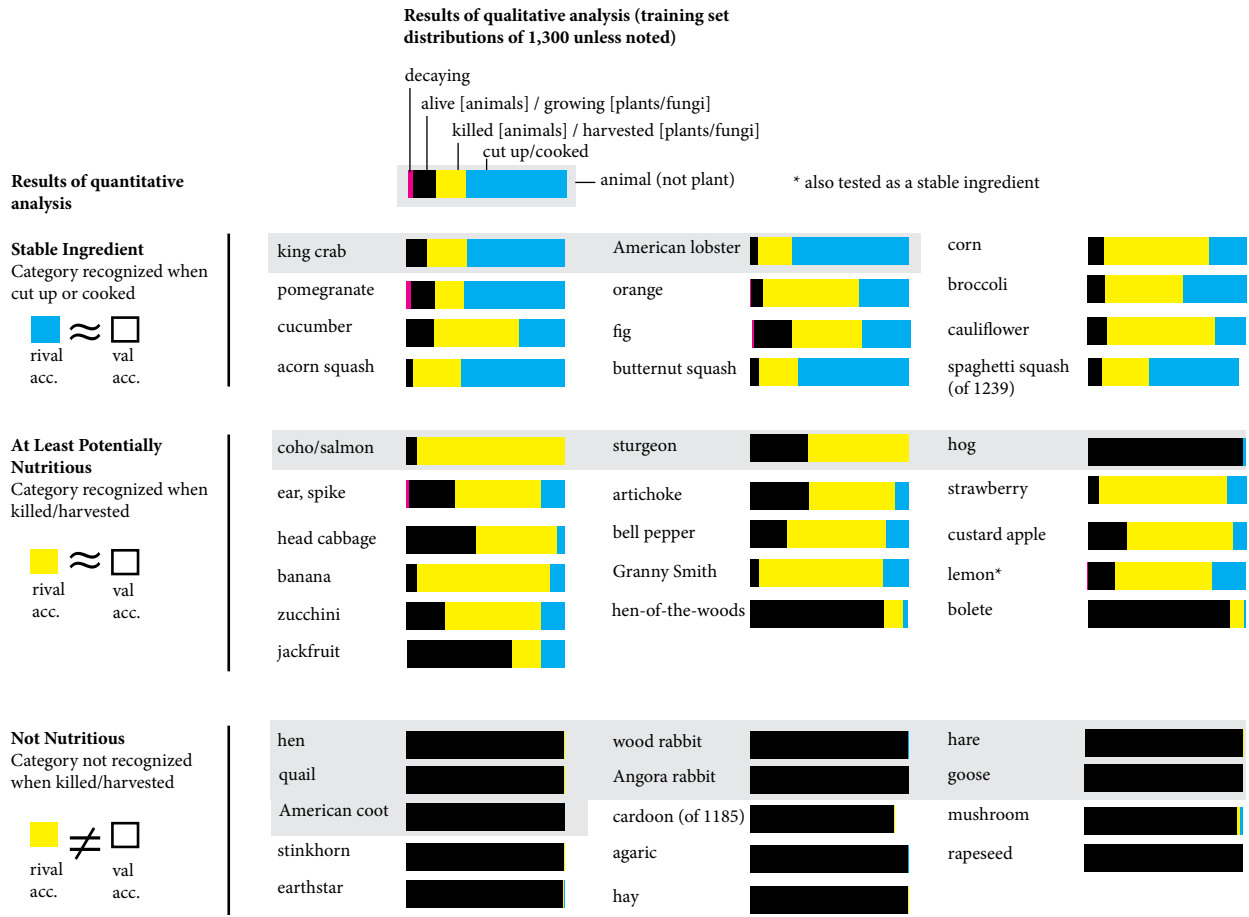

Figure 3. Detailed findings from the 42 categories in terms of their nutritiousness. Categories are sorted into three types based on what states of being VGG-16 recognizes them in. Compared to the category's validation set (val), "stable ingredients" and "potentially nutritious" categories show some similarities in their accuracy when viewed in cut up/cooked (stable) or killed/harvested (potentially nutritious) rival states (p>.1, rival sets based on 15 images of the object gathered through image searches). "Not nutritious" categories have significantly lower accuracy when viewed in killed/harvested rival state compared to the validation set (p<.01). The colored bars show that with a few surprises (hog, jackfruit, hen-of-the-woods, and bolete), the proportions of each category's training set generally predict the nutritiousness that ImageNet-trained classifiers enact (although see Section 5). Overall, ImageNet-trained classifiers are mostly pescatarian viewers, only recognizing seafood animal categories (and hogs) when they are killed. Among plant categories, ImageNet-trained classifiers view cardoons and most mushroom categories as inedible; they also don't recognize rapeseed as useful for oil or hay as edible for animals.

To the extent they're used in food applications, ImageNet-trained classifiers participate in these debates over nutrition. Perceptually, we suggest that if something is recognized when killed/harvested, it is being valued as *at least potentially nutritious*. Because cooking especially transforms things, we can ask further if something is recognized as a *stable ingredient* when cut up or cooked. Analysis of 45 categories reveals that VGG-16 is roughly a "pescatarian" (Fig. 3); among 12 commonly consumed animal categories, it acknowledges only the 4 fish and seafood categories (as well as **hog**) as themselves when they've been killed. Many common fruits and vegetables are stable ingredients or at least potentially nutritious. **Bolete** and **hen-of-the-woods** were at least potentially nutritious, but the four other edible mushroom categories (**mushroom**, **stinkhorn**, **agaric**, and **earthstar**) were not. Of 8 categories that had very few training alive/on the plant images, both **Granny Smith** apples and **spaghetti squash** went unrecognized on the plant; perceptually, these come *ready-made* at a market (not shown in Fig. 3). We were unable to obtain statistically significant results for **pineapple** (almost nutritious), **acorn** (almost not nutritious for animals), and the mushroom **coral fungus** (almost not nutritious). In consumer applications, ImageNet-trained classifiers' value on nutrition would be at odds with most US eaters, who have meat-heavy diets

[6], but would support more environmentally sustainable nutrition plans

## 4.2 Maturation

Another open question with social significance is how early in something's developmental process that thing counts as that thing. On one hand, we value *maturation* when we view something as not substantially itself until it's matured and fully developed. Certification programs, including colleges, don't award diplomas for partial completion; this intellectually and financially values students' maturation [96]. Similarly, legal expectations like an age of consent (Wikipedia) are maturation-based. Valuing maturation, then, allows us to take confidence in the functionality, preparation, and reliability of things. On the other hand, when things count as themselves even in germinal states, we value the *making-of*. This is forward-looking, focused on their potential and inclusion, and thereby also protective [93]. Valuing the making-of is particularly charged regarding human development because for many people an embryo's status as human is foundational for their legal, social, and moral stance on abortion [26][71].

Perceptually, we say a classifier values "maturation" when it doesn't recognize objects in germinal states, and that it values the "making of" when developmental states are recognized. Of 9 categories, VGG-16 strongly values maturation with living categories: none of the three frog categories (**bullfrog**, **tree frog**, **tailed frog**) were recognized as tadpoles, nor was a hybrid rival sets of various **"eggs"** (counting an image as recognized if any egg-laying animal was labeled). The exception was another hybrid category of **"seeds,"** which we generously coded as recognized when any plant category was labeled. In contrast, VGG-16 values the making-of with non-living categories: perhaps surprisingly at a philosophical level, completely untied knots (that is, just looped fabric) were recognized as **bow tie**, general **knot**, and **Windsor knot**, respectively. Similarly, uncooked pizza was still recognized as **pizza**. In educational applications, ImageNet-trained classifiers' value on the maturation of living categories rejects a "teleological essentialism" that many cultural groups hold ([63], who use tadpole-as-a-frog specifically, p.8).

## 4.3 Utility

Any mature object can be in better or worse shape; we value *utility* when we recognized things as most themselves when they are doing what they are meant to do. For humans, utility has popular expression in self-help books to find one's "purpose" [74] or "best self" [75] and in theological development to define humans relative to God [72][103]; for objects, valuing utility speaks to their productivity to us. Alternatively, seeing things as themselves even when they aren't active values *dormancy*. Valuing dormancy rejects capitalistic notions of the replaceable worker and better theorizes disability [18].

Perceptually, we identify a value on dormancy if an inactive state is recognized, and a value on utility if an inactive state isn't recognized. Of 23 categories, VGG-16 is mixed, but predominantly values utility. **Mountain tent**, (hot air) **balloon**, **prayer rug**, **torch** [flashlight], **yurt**, **parachute**, and **umbrella** are all not recognized as such when packed up. Similarly, **geyser** isn't seen when not spouting and **scuba diver** isn't recognized without water in the picture. The active states of **nail** holding material together and reimagined **triceratops** (i.e. not as a skeleton) are both recognized, affirming utility. Utility can also emerge as restricting the use of something. **Pajama** isn't recognized when it's worn outside of the house, and **stage** isn't recognized if it isn't professionally made. In contrast, VGG-16 values dormancy by recognizing **fountain** without water, **scabbard** without a sword, **bow** that isn't pulled back, **convertible** with a hard top on, **amphibious vehicle** on land without water nearby, packed up **sleeping bag**, folded **folding chair**. A **vase** with nothing in it is actually statistically easier to detect in its dormant state than vases overall. Finally, the active states of a **screw** holding material together and a **beacon** with the light on at night are both unrecognized, suggesting they are *only* visible in dormant states. In commercial applications, ImageNet-trained classifiers' value on utility would be unhelpful for storage situations in which dormancy is the goal.

## 4.4 Modesty

What counts as *modest* clothing is a large source of intergenerational and intercultural [90]. Moreover, people disagree whether immodesty is enacted by the female wearer and/or by the male viewer [67]. Debates over modesty often stand in for wider debates about gender and social roles [77].

Perceptually, we focus on items that could symbolically reveal a person's nakedness [46], although modesty is particularly complex to map. Across 12 categories, VGG-16 views somewhat modestly. It view modestly in not recognizing a **bra** that is partially covered. And because swimsuits are culturally exempted from counting as underwear, VGG-16 views modestly by identifying **bikini** and one-pieces (**maillot** and its twin **maillot, tank suit**) when they're worn with an additional top/bottom – otherwise, they would just look like underwear. In contrast, VGG-16 is not distracted by naked people; it recognizes **bath towel** and **shower cap** covering naked people and **bathtub** (and its twin **tub**) with people. **Necklace** with plunging necklines is ambiguous for modest viewers; VGG-16 doesn't recognize necklaces with plunging necklines worn by people. For consistency, we would expect similar but non-sexualized objects to be recognized, and **bathing cap** being worn around water is predictably easier to detect than bathing cap generally. But **sock** and **jersey** [t-shirt] are not recognized when covered up. This could be interpreted as incoherent or extremely modest. For personal applications, ImageNet-trained classifiers' value on modest viewing could set some users at ease.

## 4.5 Beauty

Beauty is also big business, accounting for $4.2 trillion in the U.S. [36]. By and large, this industry affirms people's *beauty efforts*. The overall message is that with the right effort and consumer savviness, anyone can make themselves beautiful, pleasing, and attractive. This carries a kind of egalitarian logic [47] (although skin-color bias lurks [48]). Rejecting a tie between consumerism and identity, people can also value *natural beauty*. This involves self-confidence in one's unadorned appearance, and negatively evaluating people's beauty efforts as vain, expensive, disempowering, and overwrought [105].

Perceptually, we say a classifier values beauty efforts if it doesn't recognize beauty items when they've been applied (i.e. they have then become part of the person); a classifier values natural beauty if it recognizes beauty items when they've been applied (a perceptual critique of their vanity). This mapping is culturally restricted by the contradictions of beauty efforts [102]. In 8 categories, VGG-16 generally values beauty efforts. It doesn't recognize **lotion**, **face powder**, **perfume**, or **hair spray** once they've been applied, although it does recognize **lipstick** on people's lips. It doesn't recognize **Band Aid** worn on someone from a distance, but it does recognize natural color **wig** when they're worn. (We didn't obtain a clear result on **sunscreen** that has been applied, even with 25 rival images.) By valuing beauty efforts, personal applications of ImageNet-trained classifiers wouldn't be useful for identifying when someone is wearing "too much" makeup.

## 4.6 Wonder

Valuing *wonder* is central to many philosophical and aesthetic approaches [31][83]. In contrast, we can also emphasize that magic tricks aren't "real," and we can break down amazing feats to show that they are within our grasp. Such a value on *mechanism* is operative especially in STEM as an intellectual agenda [5][11]. More broadly, mechanistic thinking is intimately bound up in modernity [51] and (non)religion [51][95].

Perceptually, a classifier values wonder if it doesn't recognize objects when they're "invisible" or "hidden;" it values mechanism otherwise. In 6 categories, VGG-16 is split between wonder and mechanism. It values wonder by not recognizing a disguised **walking stick** (the insect), realistic **mask**, or **jigsaw puzzle** reference image. However, it values mechanism by recognizing a camouflaged **African chameleons**, children's **backpack** shaped like a character, and **crossword** answer key. For educational purposes, ImageNet-trained classifiers' static mix of wonder/mechanism will be unsatisfying, either instantly giving away the secret or being perpetually fooled.

## 4.7 Squeamishness

Finally, it's an open question how to interact with objects in dirty/gross states. On one hand, we value *squeamishness* by exhibiting an aversion to objects with excrement, filth, bodily fluids, or graphic sexuality. Squeamishness maintains social boundaries, avoids disease, rejects ambiguity, and regulates emotional content [25][27]. On the other hand, we value *health work* when we aren't deterred by gross things. This allows certain kinds of biological knowledge [3][91] and a greater ability to recognize and critique being squeamish toward certain people, which is often a form of social bias [12].

Perceptually, we call a classifier squeamish if it doesn't recognize objects in dirty/gross states, and we say it is not squeamish or values health work if it does. Of 15 categories, VGG-16 overwhelmingly values health work. VGG-16 recognizes **toilet seat**, **diaper**, and **toilet tissue** [paper] with visible excrement, and **plunger** in toilet; **syringe** and **safety pin** piercing skin with blood; **hamper** with dirty clothes; dirty **mop** and **broom**; **garbage truck** and **ashcan** [trash can] with visible trash; and pornographic **website**. The only exceptions are **handkerchief** with visible snot or being blown into, and dirty **dishrag**. Surprisingly, our qualitative analysis of training set exceptions were very unpredictive of VGG-16's performance in this value area: toilet seat, diaper, toilet paper, plunger, mop, and website each had 10 or fewer exceptions in the training set (<0.8%), but were still recognized (plunger in toilet was actually easier to detect than plungers generally). In health applications, ImageNet-trained classifiers will be stoic and unattached when classifying objects; for some audiences, this could feel out of touch.

## 5. Findings regarding the Data Proportionality Hypothesis

The primary benefit of understanding a classifier's values is to be able to be reflective and intentional about what values a given classifier should have. This requires being able to control the values that a classifier enacts. This section draws on our extensive qualitative analysis (manual examination of over 165,000 images) and the fact that a variety of classifiers all share the same ImageNet training set. We seek to understand more precisely the extent to which exceptions in training set data predict values, and how those differ across classifier architectures. Specifically, we report results on four generations of modern classifier architectures: VGG-16 [89], ResNet50 [41], InceptionV3 [97], and NASNetLarge [107].

Our analysis is guided by three conflicting hypotheses. *1. Better classifiers learn from training set exceptions better* as part of generalizing. This would predict that a higher proportion of training set exceptions in a category would more strongly increase better classifiers' recognition of that category's rival set, even to the point of enacting a different value. *2. Better classifiers can better ignore visual context*, generalizing across scenes differently than above. This would also predict that better classifiers would recognize more rival images, even to the point of enacting a different value, but would be relatively insensitive to the proportion of training set exceptions. *3. Better classifiers are more tied to visual context*, especially since it's standard to augment training data with random image crops [58].

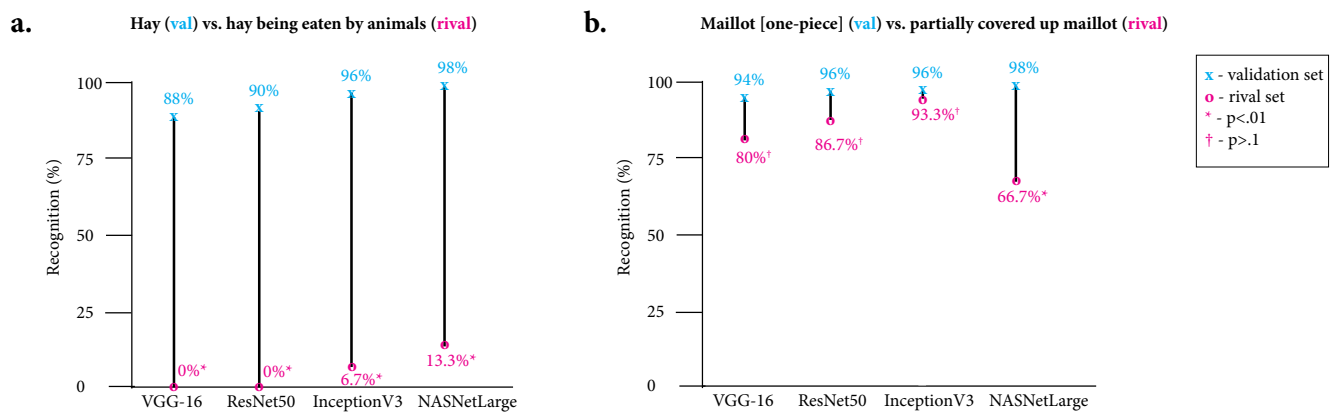

Figure 4. Comparing recognition rates across ImageNet-trained classifiers. a. A frequent pattern: hay being eaten by animals is one of 41 rival selection criteria that showed monotonic increases across classifiers on both the validation and rival sets; all classifiers value hay as not nutritious for animals (see 4.1). b. One of only 12 exceptions of 138 rival selection criteria in which value enactments differ across classifiers. VGG-16, ResNet50, and InceptionV3 view maillot modestly by recognizing them as swimwear when covered up (and thereby not as underwear), while NASNetLarge has such decreased recognition that partially covered up maillot are statistically different from maillot generally – NASNetLarge perceives maillot immodestly (see 4.4).

This would predict that better classifiers would recognize fewer of our rival images, since they are often edge cases, and would more strongly adopt the values enacted by a worse classifier.

The validation set provides a baseline accuracy for each classifier: VGG-16: 85.4%, ResNet50: 88.4%, InceptionV3: 93.0%, and NASNet: 95.6%. This progression is largely reflected on a per-category basis for the categories we examined: accuracy increases monotonically across the four classifiers on the validation set of 77 of 118 examined categories.

As far as recognition of the rival sets, **better classifiers recognized more rival images**. The averaged accuracy of all 138 rival sets is: VGG-16: 60.82%, ResNet50: 63.87%, InceptionV3: 67.42%, and NASNet: 73.31%. Moreover, 67 rival selection criteria show monotonic increases in recognition across the four classifiers, which is nearly the same as for the validation set. Of those, 41 increase monotonically on both the rival and validation sets (Fig. 4a). This supports hypotheses 1 and 2 and discounts hypothesis 3.

However, **different classifiers rarely adopted clearly different values**. Only 12 rival selection criteria showed different recognition rates so as to change the statistical assessment of the values, and not always toward greater recognition (Fig. 4b). Nutrition – agaric, earthstar, butternut squash, wood rabbit, lemon, hog. Utility – beacon, bow tie, geyser. Modesty – maillot, sock. Squeamishness - toilet tissue. These values changed in various ways. Of those 12, if we assessed the value more narrowly by comparing the top-1 recognition rates (instead of the top-5) on the validation and rival sets, only lemon, beacon, maillot, and sock would still show a clear change in value across classifiers.

Thus, all four ImageNet-trained classifiers enact mostly the same values.

To differentiate hypotheses 1 and 2 we assessed how sensitive a classifier's recognition of rival sets was to the number of exceptions it had been exposed to during training. We plotted each classifier's recognition of the rival set versus the number of images it had been exposed to with that condition in the training set, for the 99 conditions in which the exceptions formed <20% of the training set (Appendix 2). The trend line for each classifier has a positive slope, indicating that all four classifiers tended to recognize more rival set images when there were more exceptions in that category's training set. But the slope is almost identical for each classifier (Appendix 2), showing in opposition to hypothesis 1 that **better classifiers are actually not better at learning from the number of exceptions**. This suggests, with hypothesis 2, that better classifiers are better able to ignore objects' visual context.

Finally, there is dramatic variance in classifier performance for categories with extremely few exceptions, and overall **the number of exceptions don't strongly predict classifiers' performance on rival sets ($r^2$=.07-.10)**. This significantly qualifies the DPH. ML researchers cannot rely on just the absence of examples, then, when seeking to make a classifier that enacts a certain value.

## 6. Discussion

We've found that ImageNet-trained classifiers enact values in seven areas across 118 categories. Individually, these are edge cases, but collectively, value-enactment is widespread and is expected to extend to more linguistically and socially complex AI tasks. The values we've identified take a stance on significant social issues around the personal, commercial, and educational use of image classifiers. These values aren't necessarily morally wrong, but should be deliberated on and chosen consciously.

ML researchers can participate in consciously developing value-laden classifiers in several ways. First, additional datasets/classifiers, especially high-stakes and widely distributed ones, can be studied or commissioned to be studied for what values they enact. Our method in particular is meant to draw on insights from the humanities and social sciences. These studies can extend to other cultural, relational, and temporal aspects of *W*, and further examine the sources of classifiers' values. After all, datasets don't just appear, and classifier use doesn't happen in a vacuum. The results of rhetorical values analyses could drive internal and external deliberation about AI, and could parallel efforts to identify and remediate social bias in classifiers.

Second, ML researchers can develop techniques for deliberately manipulating the values that a classifier enacts. This would include examining whether values from ImageNet are "held" when used for transfer learning on a new task. Based on the findings in Section 5, this has an interesting technical challenge compared to just "fixing" social bias in classifiers; at an extreme, it could integrate with efforts for AI to approximate its social impact itself.

Finally, compared to humans who flexibly adopt various values [23][76], today's classifiers are obnoxiously static in the values they enact. ML researchers can make classifiers more lifelike by creating adaptive systems. These would use *t* strategically and could even make use of multiple classification systems, similar to the fast-reaction System 1 and slow-reaction System 2 processing that humans have evolved to use [50]. Adaptively enacting values would find particular relevance for rhetoric scholars in that an adaptive system would consider context.

## ACKNOWLEDGMENTS

The authors thank the Princeton University Committee on Research in the Humanities and Social Sciences for funding, and the Fall 2018 / Spring 2019 sections of Living with AI.

## REFERENCES

[1] A Fortiori (Accessed on Aug 22, 2019). Wikipedia.
[2] Ag and Food Sectors and the Economy (Accessed on Aug 22, 2019). USDA Economic Research Service. https://www.ers.usda.gov/data-products/ag-and-food-statistics-charting-the-essentials/ag-and-food-sectors-and-the-economy/
[3] M. M. Antony and M. A. Watling (2015). Overcoming Medical Phobias: How to Conquer Fear of Blood, Needles, Doctors, and Dentists. Echo Point Books & Media.
[4] Aristotle (2007). *On Rhetoric: A Theory of Civic Discourse*. Trans. G. A. Kennedy.
[5] R. D. Atkinson and M. Mayo (2010). Refueling the US Innovation Economy: Fresh Approaches to Science, Technology, Engineering and Mathematics (STEM) Education. ITIF. https://files.eric.ed.gov/fulltext/ED521735.pdf
[6] E. Barclay (2012). A Nation Of Meat Eaters: See How It All Adds Up. NPR. https://www.npr.org/sections/thesalt/2012/06/27/155527365/visualizing-a-nation-of-meat-eaters
[7] B. Berlin and P. Kay (1969). *Basic Color Terms: Their Universality and Evolution*. U California Press.
[8] "Brassica oleracea" (Accessed on Aug 22, 2019). Wikipedia. https://en.wikipedia.org/wiki/Brassica_oleracea
[9] J. S. Bruner and C. C. Goodman (1947). Value and Need as Organizing Factors in Perception. The Journal of Abnormal and Social Psychology, 42(1), 33-44.
[10] J. Buolamwini and T. Gebru (2018). Gender Shades: Intersectional Accuracy Disparities in Commercial Gender Classification. FAT*.
[11] P. Carnevalle, N. Smith, M. Melton (2011). STEM: Science, Technology, Engineering, Mathematics. Center on Education and the Workforce. https://files.eric.ed.gov/fulltext/ED525297.pdf
[12] J. D. Cisneros (2008). Contaminated Communities: The Metaphor of "Immigrant as Pollution" in Media Representations of Immigration. Rhetoric & Public Affairs, 11(4), 569-602.
[13] Clark and D. Chalmers (1997). The Extended Mind. Analysis, 58(1), 7-19.
[14] M. C. Coleman (2018). Machinic Rhetorics and the Influential Movements of Robots. Review of Communication 18(4), 336-351.
[15] Community Standards Enforcement Report (2019). Facebook Transparency. https://transparency.facebook.com/community-standards-enforcement#bullying-and-harassment
[16] E. D. Cubuk et al. (2018). AutoAugment: Learning Augmentation Policies from Data. CVPR.
[17] D. Davis (2010). Inessential Solidarity: Rhetoric and Foreigner Relations. University of Pittsburgh Press.
[18] L. J. Davis (2002). Bodies of difference: Politics, disability, and representation. In *Disability Studies: Enabling the Humanities*, ed. S. L. Snyder, B. J. Brueggemann, and R. Garland-Thomson.
[19] T. DeVries, I. Misra, C. Wang, and L. van der Maaten (2019). Does Object Recognition Work for Everyone? CVPR.
[20] T. Demo (Ed.) (2015). *Rhetoric Across Borders*. Parlor Press, Anderson, South Carolina.
[21] J. Deng, et al. (2009). ImageNet: A Large-Scale Hierarchical Image Database. CVPR.
[22] M. R. Dickey (2017). Algorithmic Accountability. TechCrunch Apr, 30. https://techcrunch.com/2017/04/30/algorithmic-accountability/
[23] P. DiMaggio (1997). Culture and Cognition. Annual Review of Sociology, 23, pp.263-287.
[24] T. Doshi (2018). Introducing the Inclusive Images Competition. Google AI Blog, Sept. 6. https://ai.googleblog.com/2018/09/introducing-inclusive-images-competition.html
[25] M. Douglas (1966). Purity and Danger: An Analysis of the Concepts of Pollution and Taboo. Routledge, London and New York.
[26] N. Edgar (2017). The Rhetoric of Auscultation: Corporeal Sounds, Mediated Bodies, and Abortion Rights. Quarterly Journal of Speech, 103(4), 350-371.
[27] M. H. Erdelyi (1974). A New Look at the New Look: Perceptual Defense and Vigilance. Psychological Review, 81(1), 1-25.
[28] L. Fei Fei and J. Deng (2017). ImageNet: Where Have We Been? Where Are We Going? ILSVRC presentation.
[29] K. Ferryman and M. Pitcan (2018). Fairness in Precision Medicine. Data & Society, Feb. https://datasociety.net/wp-content/uploads/2018/02/Data.Society.Fairness.In_.Precision.Medicine.Feb2018.FINAL-2.26.18.pdf
[30] R. Field (Accessed on Aug. 22, 2019). John Dewey (1859-1952). Internet Encyclopedia of Philosophy. A Peer-Reviewed Academic Resource. https://www.iep.utm.edu/dewey/
[31] P. Fisher (2003). Wonder, the Rainbow, and the Aesthetics of Rare Experiences. Harvard University Press.
[32] L. Flower, E. Long and L. Higgins. *Learning to Rival: A Literate Practice for Intercultural Inquiry*. Lawrence Erlbaum Associates, Mahwah, New Jersey.
[33] B. Friedman and H. Nissenbaum (1996). Bias in Computer Systems. ACM Transactions on Information Science 14(3), 330-347.
[34] D. Gershgorn (2017). The data that transformed AI research—and possibly the world. Quartz, July 16. https://qz.com/1034972/the-data-that-changed-the-direction-of-ai-research-and-possibly-the-world/
[35] J. J. Gibson (1979). The Ecological Approach to Visual Perception. Routledge.
[36] 2018 Global Wellness Economy Monitor (2018). Global Wellness Institute. https://globalwellnessinstitute.org/industry-research/2018-global-wellness-economy-monitor/
[37] D. Goodman and M. Redcliffe (2002). *Refashioning Nature: Food, Ecology, and Culture*. Routledge.
[38] D. Graham (2015). Heraclitus. Stanford Encyclopedia of Philosophy. https://plato.stanford.edu/entries/heraclitus/
[39] N. Hannah-Jones (2019). Our democracy's founding ideals were false when they were written. Black Americans have fought to make them true. New York Times Magazine, Aug 14. https://www.nytimes.com/interactive/2019/08/14/magazine/black-history-american-democracy.html

[40] G. A. Hauser (1999). Vernacular Voices: The Rhetoric of Publics and Public Spheres. University of South Carolina.

[41] K. He, X. Zhang, S. Ren, and J. Sun (2015). Delving Deep into Rectifiers: Surpassing Human-Level Performance on ImageNet Classification. ICCV

[42] M. Heidegger (1996 [1927]). *Being and Time*, trans. J. Stambaugh. State University of New York Press.

[43] J. Henrich (2008). A Cultural Species: Why a Theory of Culture Required to Build a Science of Human Behavior. In *Explaining Culture Scientifically*, M. Brown (Ed.) (pp. 184-210). University of Washington Press.

[44] J. Henrich, S. Heine, A. Norenzayan (2010). The Weirdest People in the World? Behavior and Brain Sciences 33, 61-135.

[45] W. S. Hesfod, A. C. Licona, and C. Teston (Eds.) *Precarious Rhetorics*. The Ohio State University Press, Columbus.

[46] Hollander (1993 [1975]). *Seeing Through Clothes*. University of California Press.

[47] R. Holliday and J. C. Taylor (2006). Aesthetic Surgery as False Beauty. Feminist Theory, 7(2), 179-195.

[48] M. L. Hunter (2011). Buying Racial Capital: Skin-Bleaching and Cosmetic Surgery in a Globalized World. Journal of Pan-African Studies, 4(4), 142-164.

[49] J. Johnston and S. Baumann (2014). Foodies: Democracy and Distinction in the Gourmet Foodscape. Routledge.

[50] D. Kahneman (2013). *Thinking, Fast and Slow*. Farrar, Straus and Giroux

[51] M. Kang (2011). Sublime Dreams of Living Machines: The Automaton in the European Imagination. Harvard University Press.

[52] D. Kaplan (Ed.) (2012). *The Philosophy of Food*. University of California Press.

[53] I. X. Kendi (2016). Stamped from the Beginning: The Definitive History of Racist Ideas in America. Nation Books, New York.

[54] Using Pre-Trained Models (Accessed Aug 22, 2019). Keras documentation. https://keras.rstudio.com/articles/applications.html

[55] Khosla, T. Zhou, T. Malisiewicz, A. A. Efros, and A. Torralba (2012). Undoing the Damage of Dataset Bias. ECCV.

[56] S. Kitayama, S. Duffy, T. Kawamura, and J. T. Larsen (2003). Perceiving an Object and Its Context in Different Cultures: A Cultural Look at New Look. Psychological Science 14(3), 201-206.

[57] S. Krishnapriya, K. Vangara, M. King, V. Albiero, and K. Bowyer (2019). Characterizing the Variability in Face Recognition Accuracy Relative to Race. CVPR.

[58] Krizhevsky, I. Sutskever, and G. Hinton (2013). ImageNet Classification with Deep Convolutional Neural Networks. NIPS.

[59] T. Kuhn (1962). The Structure of Scientific Revolutions. University of Chicago Press.

[60] Y. LeCun, L. Bottou, Y. Bengio, and P. Haffner (1988). Gradient-Based Learning Applied to Document Recognition. IEEE.

[61] Y. LeCun et al. (1989). Backpropagation Applied to Handwritten Zip Code Recognition. Neural Computation 1, 541-551.

[62] Y. LeCun, Y. Bengio, and G. Hinton (2015). Deep Learning. Nature, 521, pp.436-444.

[63] D. Medin and S. Atran (2004). The Native Mind: Biological Categorization and Reasoning in Development and Across Cultures. Psychological Review.

[64] K. H. F. Meyer and R. Land (2003). Threshold Concepts and Troublesome Knowledge – Linkages to Ways of Thinking and Practising. In *Improving Student Learning – Ten Years On*. C. Rust (Ed), OCSLD, Oxford

[65] W. Mignolo (2011). The Darker Side of Western Modernity: Global Futures, Decolonial Options. Duke University Press.

[66] C. M. Millward *A Biography of the English Language* (2nd ed.). Thomson Wadsworth.

[67] L. Mulvey (1975). Visual Pleasure and Narrative Cinema. Screen, 16(3), 6–18.

[68] R. E. Nisbett K. Peng, I. Choi, and A. Norenzayan (2001). Culture and Systems of Thought: Holistic Versus Analytic Cognition. Psychological Review 108(2), 291-310.

[69] R. E. Nisbett and Y. Miyamoto (2005). The Influence of Culture: Holistic Versus Analytic Perception. Trends in Cognitive Science, 9(10), 468-473.

[70] S. U. Noble (2018). Algorithms of Oppression: How Search Engines Reinforce Racism. New York University Press.

[71] J. T. Noonan, Jr. (1970). The Morality of Abortion: Legal and Historical Perspectives. Harvard University Press.

[72] O. O'Donovan (1980). The Problem of Self-Love in St. Augustine. Wipf & Stock Pub.

[73] b. l. ojalehto and D. L. Medin (2015). Perspectives on Culture and Concepts. Annual Review of Psychology 66, 249-275.

[74] Oprah (2019). The Path Made Clear: Discovering Your Life's Direction and Purpose. Flatiron Books.

[75] J. Osteen (2004). Your Best Life Now: 7 Steps to Living at Your Full Potential. FaithWords.

[76] O. Patterson (2014). Making Sense of Culture. Annual Review of Sociology. 40, 1-30.

[77] P. Pender (2012). Early Modern Women's Writing and the Rhetoric of Modesty. Palgrave Macmillan.

[78] Peskoe-Yang (2019). Vegetables Don't Exist. Popula, Feb. 20. https://popula.com/2019/02/20/vegetables-dont-exist

[79] D. Raji and J. Buolamwini (2019). Actionable Auditing: Investigating the Impact of Publicly Naming Biased Performance Results of Commercial AI Products. AIES.

[80] T. Regier and P. Kay (2009). Language, Thought, and Color: Whorf Was Half Right. Trends in Cognitive Science, 13(10), 439-446.

[81] S. Roose (2009). The Making of a YouTube Radical. New York Times Magazine, June 8. https://www.nytimes.com/interactive/2019/06/08/technology/youtube-radical.html

[82] G. Rose (2003). Visual Methodologies: An Introduction to Researching with Visual Materials (4th ed.). Sage.

[83] M. Rubenstein (2008). Strange Wonder: The Closure of Metaphysics and the Opening of Awe. Columbia University Press.

[84] O. Russakovsky et al. (2014). ImageNet Large Scale Visual Recognition Challenge. IJCV.

[85] I. J. Ryan, N. Myers and R. Jones (2016). *Rethinking Ethos: A Feminist Ecological Approach to Rhetoric*. Southern Illinois University Press.

[86] D. Sedley (1982). The Stoic Criterion of Identity. Phronesis, 27(3), 255-275.

[87] I. H. Segall, D. T. Campbell, and M. J. Herskovits (1966). *The Influence of Culture on Visual Perception*. Bobbs-Merrill Co.

[88] T. Simonite (2018). When It Comes to Gorillas, Google Photos Remains Blind. Wired, Jan 11. https://www.wired.com/story/when-it-comes-to-gorillas-google-photos-remains-blind/

[89] K. Simonyan and A. Zisserman (2015). Very Deep Convolutional Networks for Large-Scale Image Recognition. ICLR.

[90] R. Sobh, R. W. Belk and J. Gressel (2012). Modest Seductiveness: Reconciling Modesty and Vanity by Reverse Assimilation and Double Resistance. Journal of Consumer Behavior 11, 357-367.

[91] D. Solot and A. Arluke (1997). Learning the Scientist's Role: Animal Dissection in Middle School. Journal of Contemporary Ethnography, 26(1), 28-54.

[92] G. Sonnante, D. Pignone, and K. Hammer (2007). The Domestication of Artichoke and Cardoon: From Roman Times to the Genomic Age. Annals of Botany 100, 1095-1100.

[93] J. G. Speth (2008). *The Bridge at the Edge of the World: Capitalism, the Environment, and Crossing from Crisis to Sustainability*. Yale University Press.

[94] L. Stark (2019). Facial Recognition is the Plutonium of AI. XRDS, 25(3), 50-55.

[95] V. Stenger (2007). God The Failed Hypothesis. How Science Shows that God Does Not Exist. Prometheus.

[96] V. Tinto (2012). Completing College: Rethinking Institutional Action. University of Chicago Press.

[97] C. Szegedy, V. Vanhoucke, S. Ioffe, J. Shlens, Z. Wojna (2016). Rethinking the Inception Architecture for Computer Vision. CVPR.

[98] Torralba and A. A. Efros (2011). Unbiased Look at Dataset Bias. CVPR.

[99] US Weight Loss and Diet Control Market (2019). Marketdata LLC, February. https://www.researchandmarkets.com/research/qm2gts/the_72_billion?w=4

[100] L. Walsh et al. (2017). Forum: Bruno Latour on Rhetoric. Rhetoric Society Quarterly, 47:5, 403-462.

[101] D. A. Wanzer[-Serrano] (2012). Delinking Rhetoric, or Revisiting McGee's Fragmentation Thesis through Decoloniality. Rhetoric and Public Affairs 15(4), 647-657.

[102] B. R. Weber (2009). Makeover TV: Selfhood, Citizenship, and Celebrity. Duke University Press.

[103] Westminster Shorter Catechism (Accessed on Aug. 22 2019 [1647]). Shorter Catechism of the Assembly of Divines. http://www.apuritansmind.com/westminster-standards/shorter-catechism/

[104] J. K. Witt (2001). Action's Effect on Perception. Current Directions in Psychological Science, 20(3), 201-206.

[105] L. Wolf (1991). The Beauty Myth: How Images of Beauty are Used Against Women. Harper-Collins.

[106] Y. J. Xiao, G. Coppin and J. J. Van Bavel (2016). Perceiving the World Through Group-Colored Glasses: A Perceptual Model of Intergroup Relations. Psychological Inquiry, 27(4), 255-274.

[107] B. Zoph, V. Vasudevan, J. Shlens, Q. V. Le (2017). Learning Transferable Architectures for Scalable Image Recognition. CVPR.

# Appendix 1

Detailed view of the categories we examined and the comparisons we made, with commentary below.
Similarity in p-value:
"Extremely low" p ≤.0001
"Low" .0001 < p < .01
"Unclear" .01 ≤ p < .1
"High" .1 < p < .5
"Extremely high" .5 ≤ p
"Easier to detect" p ≤.01 (where the rival recognition rate is higher than the validation recognition rate)

| Value area and value enacted | Category - rival criteria | % rival (VGG-16 top-5) | % val (VGG-16 top-5) | Similarity (p-value) |
|---|---|---|---|---|
| NUTRITION | | | | |
| Stable ingredient | **king crab, Alaska crab, Alaskan king crab, Alaska king crab, Paralithodes camtschatica** - cut up | 100% | 94% | Extremely high |
| | **American lobster, Northern lobster, Maine lobster, Homarus americanus** - cut up | 100% | 88% | High |
| | **corn** - cut up | 80% | 86% | Extremely high |
| | **pomegranate** - cut up | 100% | 94% | Extremely high |
| | **orange** - cut up | 93% | 90% | Extremely high |
| | **broccoli** - cut up | 100% | 96% | Extremely high |
| | **cucumber, cuke** - cut up | 100% | 96% | Extremely high |
| | **fig** - cut up | 80% | 88% | High |
| | **cauliflower** - cut up | 80% (of 25) | 94% | High |
| | **acorn squash** - cut up | 100% | 94% | Extremely high |
| | **butternut squash** - cut up | 53% | 74% | High |
| | **spaghetti squash** - cut up | 93% | 92% | Extremely high |
| At least potentially nutritious | **coho, cohoe, coho salmon, blue jack, silver salmon, Oncorhynchus kisutch** - killed | 100% | 88% | High |
| | **sturgeon** - killed | 60% (of 20) | 80% | High |
| | **hog, pig, grunter, squealer, Sus scrofa** - killed | 60% | 76% | High |
| | **ear, spike, capitulum** - harvested | 100% | 88% | High |
| | **artichoke, globe artichoke** - harvested | 100% | 90% | Extremely high |
| | **strawberry** - harvested | 93% | 84% | Extremely high |
| | **head cabbage** - harvested | 100% | 92% | Extremely high |
| | **bell pepper** - harvested | 87% | 94% | High |
| | **custard apple** - harvested | 92% (of 14) | 90% | Extremely high |
| | **banana** - harvested | 87% | 88% | Extremely high |
| | **Granny Smith** - cut up | 87% | 96% | High |
| | **lemon** - harvested | 100% | 92% | Extremely high |
| | **zucchini, courgette** - harvested | 100% | 90% | Extremely high |
| | **acorn** - harvested | 71% (of 14) | 96% | Unclear |
| | **pineapple, ananas** - harvested | 100% (of 25) | 86% | Unclear |
| | **jackfruit, jak, jack** - harvested | 100% | 100% | Extremely high |
| | **hen-of-the-woods, hen of the woods, Polyporus frondosus, Grifola frondosa** - harvested | 93% | 92% | Extremely high |
| | **bolete** - harvested | 87% | 86% | Extremely high |
| Not Nutritious | **hen** - killed | 60% (of 20) | 94% | Low |
| | **wood rabbit, cottontail, cottontail rabbit** - killed | 47% | 94% | Low |
| | **hare** - killed | 0% | 98% | Extremely low |
| | **quail** - killed | 0% | 96% | Extremely low |
| | **Angora, Angora rabbit** - killed | 0% | 100% | Extremely low |
| | **goose** - killed | 20% | 82% | Extremely low |
| | **American coot, marsh hen, mud hen, water hen, Fulica americana** - killed | 7% | 96% | Extremely low |
| | **cardoon** - harvested | 7% | 94% | Extremely low |

| Value area and value enacted | Category - rival criteria | % rival (VGG-16 top-5) | % val (VGG-16 top-5) | Similarity (p-value) |
|---|---|---|---|---|
| | mushroom - harvested | 20% | 96% | Extremely low |
| | stinkhorn, carrion fungus - harvested | 56% (of 16) | 96% | Low |
| | agaric - harvested | 60% | 92% | Low |
| | rapeseed - in oil form | 0% | 96% | Extremely low |
| | earthstar - harvested | 67% | 100% | Extremely low |
| | coral fungus - harvested | 80% (of 20) | 98% | Unclear |
| | hay - being eated by animals | 0% | 88% | Extremely low |
| MATURATION | | | | |
| Maturation | bullfrog, Rana catesbeiana - tadpoles | 20% | 86% | Extremely low |
| | tree frog, tree-frog - tadpoles | 7% | 90% | Extremely low |
| | tailed frog, bell toad, ribbed toad, tailed toad, Ascaphus trui - tadpoles | 7% | 66% | Extremely low |
| | eggs - variety of eggs [any bird category counts as correct] | 7% | N/A | N/A |
| Making-of | seeds - variety of seeds [any plant category counts as correct] | 80% | N/A | N/A |
| | bow tie, bow-tie, bowtie - fully untied | 73% | 84% | High |
| | knot - fully untied | 93% | 80% | High |
| | Windsor tie - fully untied | 60% | 56% | Extremely high |
| | pizza, pizza pie - uncooked | 93% | 84% | Extremely high |
| UTILITY | | | | |
| Utility | mountain tent - fully packaged | 0% | 94% | Extremely Low |
| | balloon - being packaged | 40% | 96% | Extremely Low |
| | prayer rug, prayer mat - completely rolled up | 20% | 86% | Extremely Low |
| | torch - unlit | 20% | 90% | Extremely Low |
| | yurt - frame folded up | 13% | 98% | Extremely Low |
| | parachute, chute - in backpack completely compacted | 50% (of 16) | 94% | Low |
| | umbrella - closed | 20% (of 20) | 66% | Low |
| | geyser - dormant | 73% | 98% | Low |
| | scuba diver - water not visible | 47% | 96% | Extremely Low |
| | nail - in use | 67% | 76% | Extremely High |
| | triceratops - alive | 100% | 88% | High |
| | pajama, pyjama, pj's, jammies - outside of a house | 40% | 78% | Low |
| | stage - impromptu | 7% | 88% | Extremely Low |
| Dormancy | fountain - turned off | 80% (of 20) | 92% | High |
| | scabbard - without sword | 93% | 76% | High |
| | bow - not being pulled back | 93% | 82% | High |
| | convertible - hard top on | 93% | 94% | Extremely High |
| | amphibian, amphibious vehicle - on land without water visible | 87% | 90% | Extremely High |
| | sleeping bag - rolled up in package | 73% | 76% | Extremely High |
| | folding chair - folded up | 60% | 72% | Extremely High |
| | vase - without anything in it | 100% | 64% | Easier to detect |
| | screw - in use | 33% | 100% | Extremely Low |
| | beacon - light on at night | 73% | 98% | Low |
| MODESTY | | | | |
| Modest Viewing | brassiere, bra, bandeau - at least partly covered | 13% | 68% | Low |
| | bikini, two-piece - covered up | 79% (of 14) | 78% | Extremely High |
| | maillot - top or bottom worn on top | 80% | 94% | High |

| Value area and value enacted | Category - rival criteria | % rival (VGG-16 top-5) | % val (VGG-16 top-5) | Similarity (p-value) |
|---|---|---|---|---|
| | maillot, tank suit - top or bottom worn on top | 73% | 82% | High |
| Immodest Viewing | bath towel - covering naked person | 60% | 76% | High |
| | shower cap - worn in a bathroom context | 87% | 78% | Extremely High |
| | bathtub, bathing tub, bath, tub - with nude adult | 100% (of 20) | 78% | Easier to detect |
| | tub, vat - with nude adult | 93% | 74% | High |
| Unclear | necklace - revealing | 27% | 98% | Extremely Low |
| | bathing cap, swimming cap - worn around a pool/body of water | 100% | 80% | Easier to detect |
| | sock - shoe worn on top, and/or pant/skirt/dress bottom covering the top up | 27% | 74% | Low |
| | jersey, T-shirt, tee shirt - partly covered by a top | 0% | 90% | Extremely Low |
| BEAUTY | | | | |
| Beauty efforts | lotion - not in container | 33% | 84% | Low |
| | face powder - no package or applicator visible | 0% | 78% | Extremely low |
| | perfume - no container visible | 13% | 88% | Extremely low |
| | hair spray - no container visible | 28% (of 25) | 66% | Low |
| | Band Aid - medium to far distance | 13% | 76% | Extremely low |
| Natural beauty | lipstick, lip rouge - worn by someone no container visible middle/far distance | 60% | 74% | High |
| | wig - naturalistic colors on a person | 93% | 84% | Extremely high |
| Unclear | sunscreen, sunblock, sun blocker - at least mostly rubbed in, no container visible | 32% (of 25) | 60% | Unclear |
| | - | | | |
| WONDER | - | | | |
| Wonder | walking stick, walkingstick, stick insect - very blended in | 33% | 86% | Low |
| | mask - realistic [natural skin tones including celebrity] | 13% | 70% | Low |
| | jigsaw puzzle - reference image or contructed puzzle | 40% | 92% | Extremely low |
| Mechanism | African chameleon, Chamaeleo chamaeleon - very blended in | 80% | 84% | Extremely high |
| | backpack, back pack, knapsack, packsack, rucksack, haversack - has a character on the front and is shaped that way | 67% | 54% | Extremely high |
| | crossword puzzle, crossword - answer key or constructed puzzle | 80% | 80% | Extremely high |
| SQUEAMISHNESS | | | | |
| Health work | toilet seat - visible excrement | 80% | 80% | Extremely high |
| | diaper, nappy, napkin - visible excrement | 73% | 76% | Extremely high |
| | toilet tissue, toilet paper, bathroom tissue - visible excrement | 53% | 68% | High |
| | syringe - piercing skin with blood visible | 73% | 64% | Extremely high |
| | safety pin - stuck in skin, drawing blood, piercing someone | 47% | 60% | High |
| | web site, website, internet site, site - pornography | 100% | 100% | Extremely high |
| | paper towel - visibly dirty | 65% (of 20) | 82% | High |
| | plunger, plumber's helper - with dirty toilet | 87% | 52% | Easier to detect |
| | broom - visibly dirty | 67% | 74% | Extremely high |
| | swab, swob, mop - visibly dirty | 93% | 90% | Extremely high |
| | hamper - with visibly dirty laundry | 80% | 80% | Extremely high |
| | garbage truck, dustcart - garbage visible in the back | 100% | 100% | Extremely high |
| | ashcan, trash can, garbage can, wastebin, ash bin, ash-bin, ashbin, dustbin, trash barrel, trash bin - garbage visible | 53% | 74% | High |
| Squeamishness | handkerchief, hankie, hanky, hankey - visible snot/mucus or being used by someone to blow nose | 13% | 90% | Extremely low |
| | dishrag, dishcloth - visibly dirty | 7% | 94% | Extremely low |

Commentary on Appendix 1.

For all comparisons, we use ImageNet's public 50,000-image validation set in lieu of the testing set [41] [58].

We view our analysis of these shapeshifting categories as comprehensive but not exhaustive. Rather, we make use of *a fortiori* analytic decisions [1] by limiting our analytic effort when we expect that the results would be even more extreme in the same cultural context. For instance, since even chickens are not recognized when killed, we don't analyze the various dog categories to see if they are recognizable when being prepared to be eaten.

For clarity, we analyzed each category with regard to a single value, but the boundaries of a single category could easily speak to multiple values. For instance, agaric is a mushroom category that we analyze in terms of its nutritiousness. But agaric is also hallucinogenic and illegal in many areas, so it could also be analyzed as speaking to a value on legal authority.

When possible, we assessed classifiers' enacted values consistently across objects. Thus, for nutrition categories, we divided all plants and fungi between "on the plant" and "harvested" (and divided all animals between "alive" and "killed"). This provides consistency but does simplify variation across categories. For instance, quail are killed early in the food production cycle whereas crab and lobster are not. Likewise, some mushroom categories were seen on the ground partly broken from the stem, making them hard to identify as either on the plant or harvested.

Based on preliminary tests, we found that 15 rival images would usually provide sufficient distinction; in rare cases when the results were ambiguous (.01 < p < .1), we subsequently added 5 more rival images.

Procedurally, an open question relevant to one category often snowballed to other categories. The open question of object activation, for instance, applies not just to parachutes but to the ImageNet categories "mountain tent," "balloon," "folding chair," "umbrella," and others. Thus, our heuristic for characterizing dataset variation sometimes led the way to identifying an open question, and sometimes followed.

We made further comparisons not reported in this table that functioned as preliminary.

## Appendix 2

Comparison of how the training set exceptions (manually counted) relate to different classifiers' performance. Recognition rates are striated because we generally gathered 15 images for each rival set; each horizontal line represents another image correct in that rival set.

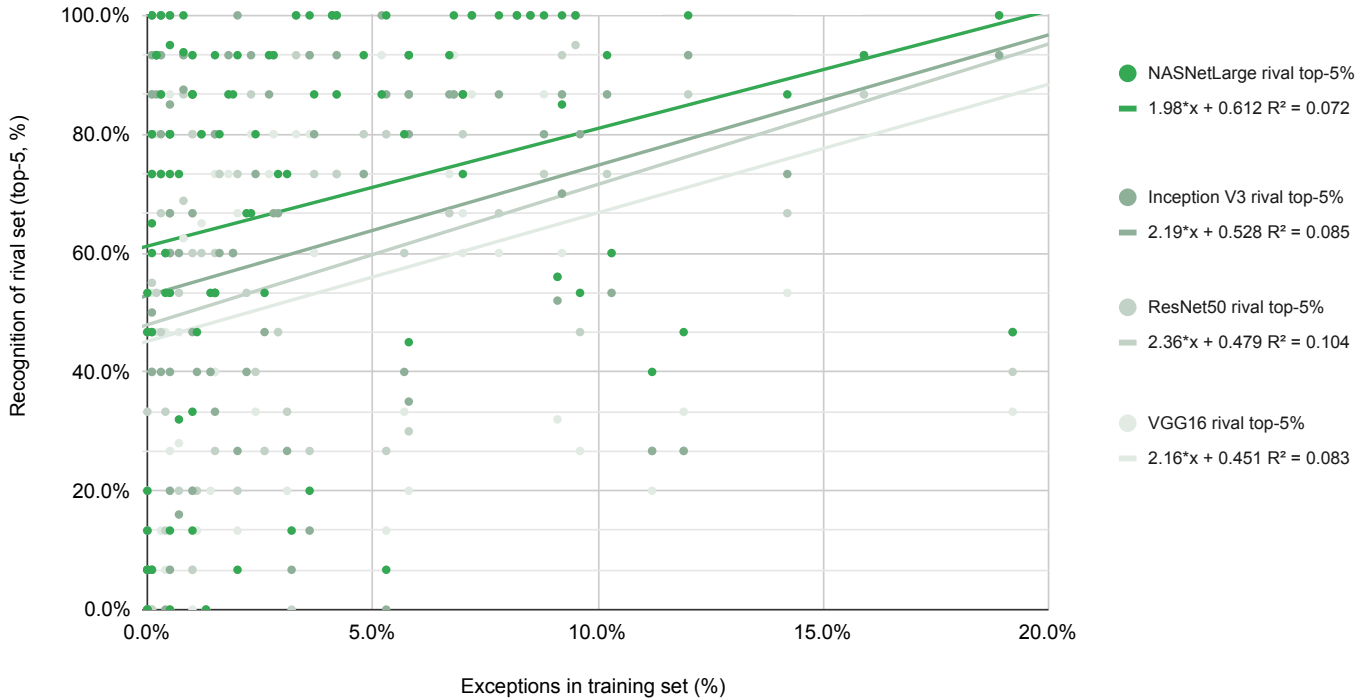

See Section 5. As an example, along the NASNetLarge trend line is a point for lipstick, which it had been exposed to in the training set 6.8% when worn by someone with no container visible. Further along the trend line is ashcans, which NASNetLarge had been exposed to with visible trash in 14.2% of the training set. Correspondingly, NASNetLarge recognized more ashcans with visible trash (86.7%) than it did lipstick worn by people with no container (73.3%). In those cases, the number of training set exceptions predicted the recognition rates well. However, the other data points show that recognition rates are quite varied for all classifiers on categories throughout, especially those with extremely few exceptions (5% = 65 training images). Overall, only 7-10% of the variation in recognition is explained statistically by the number of exceptions a classifier had been exposed to.

# Appendix 3

A sample rival set: "partially hidden sock." We used image search tools to obtain a variety of images with our criteria. We numbered the images in the order we obtained them and gave each one descriptive names.

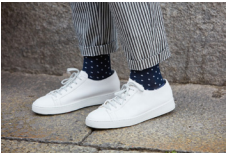

1 white on navy blue with white shoes and striped pants close distance on stone

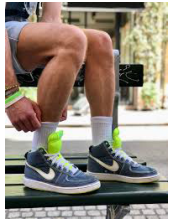

2 white long with blue nike sneakers close distance on park bench

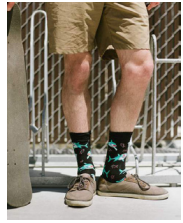

3 black with blue sharks and beige shoes close distance on pavement

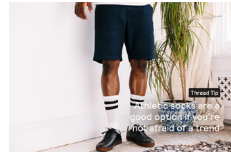

4 white with two black stripes on top with black shoes close distance in white room

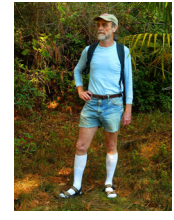

5 white with sandals middle/far distance in the woods

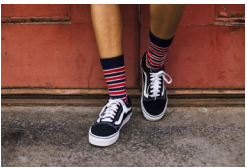

6 red white and blue striped with vans close distance in front of red door

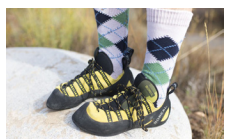

7 white with green diamonds with hiking shoes close distance on white rock

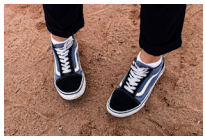

8 white ankle socks black and blue vans close distance on dirt

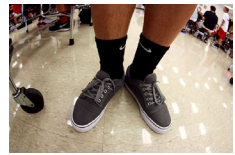

9 black with white nike symbol with gray shows close distance in store

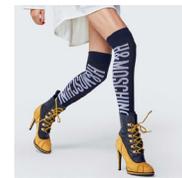

10 black with white print moschino socks with heeled boots close distance white background

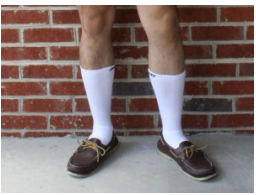

11 white with black nike symbol with boat shoes close distance in front of brick wall

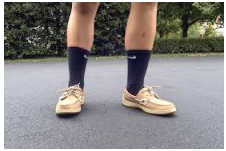

12 black with white nike symbol and boat shoes close distance on driveway

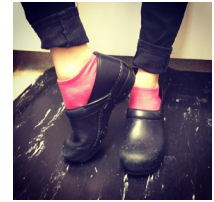

13 pink with clogging shoesclose distance on black floor

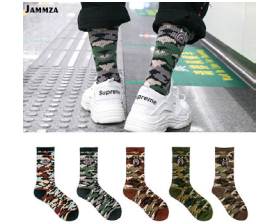

14 camo print with white supreme shows close distance on gray ground

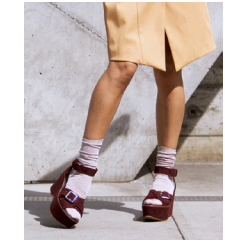

15 white with heels close distance on pavement

Top-5 recognition results:

| Image | VGG-16 | ResNet50 | InceptionV3 | NASNetLarge |
|---|---|---|---|---|
| 1 | ✓ | | ✓ | ✓ |
| 2 | | | | |
| 3 | | | ✓ | |
| 4 | | | ✓ | ✓ |
| 5 | | | ✓ | |
| 6 | | ✓ | ✓ | ✓ |
| 7 | ✓ | ✓ | ✓ | ✓ |
| 8 | | ✓ | | |
| 9 | | ✓ | ✓ | |
| 10 | | ✓ | ✓ | ✓ |
| 11 | | | ✓ | ✓ |
| 12 | | | ✓ | |
| 13 | ✓ | ✓ | ✓ | ✓ |
| 14 | ✓ | ✓ | ✓ | ✓ |
| 15 | | | | |

Sock was one of twelve categories that "flipped" values across classifiers. For VGG-16, there's significantly less recognition of partially hidden socks (4/15≈27%) than of socks generally (37/50=74%), p=.002. NASNetLarge is also significantly different from its validation recognition rate (p=.004), even though it has double the rival recognition rate of VGG-16; ResNet50 is indeterminate by our standards (p=.05). For InceptionV3, however, there's high similarity between its recognition on the hidden socks (12/15=80%) and its recognition on socks generally (44/50=88%), p=.41. Of the 1,300 image training set, 9.6% (125) images showed socks that were partially hidden.